\documentclass[graybox]{svmult}

\usepackage{mathptmx}       
\usepackage{helvet}         
\usepackage{courier}        
\usepackage{type1cm}        
\usepackage{makeidx}         
\usepackage{graphicx}        
\usepackage{multicol}        
\usepackage[bottom]{footmisc}

\makeindex             

\begin{document}

\title*{Disc formation in turbulent cloud cores: Circumventing the magnetic braking catastrophe}
\titlerunning{Disc formation in turbulent cloud cores}
\author{D. Seifried, R. Banerjee, R.~E. Pudritz and R.~S. Klessen}
\institute{D. Seifried \at Hamburger Sternwarte, University of Hamburg, Gojenbergsweg 112, 21029 Hamburg \email{dseifried@hs.uni-hamburg.de}}
%
%
\maketitle

\abstract{We present collapse simulations of strongly magnetised, 100 M$_{\odot}$, turbulent cloud cores. Around the protostars formed during the collapse Keplerian discs with typical sizes of up to 100 AU build up in contrast to previous simulations neglecting turbulence. Analysing the condensations in which the discs form, we show that the magnetic flux loss is not sufficient to explain the build-up of Keplerian discs. The average magnetic field is strongly inclined to the disc which might reduce the magnetic braking efficiency. However, the main reason for the reduced magnetic braking efficiency is the highly disordered magnetic field in the surroundings of the discs. Furthermore, due to the lack of a coherently rotating structure in the turbulent environment of the disc no toroidal magnetic field necessary for angular momentum extraction can build up. Simultaneously the angular momentum inflow remains high due to local shear flows created by the turbulent motions. We suggest that the "magnetic braking catastrophe" is an artefact of the idealised non-turbulent initial conditions and that turbulence provides a natural mechanism to circumvent this problem.}

\section{Introduction}
\label{sec:intro}

In recent years a number of authors have studied the formation of protostellar discs during the collapse of strongly magnetised magnetic cloud cores~\cite{Hennebelle08a,Seifried11}. In simulations with magnetic field strengths comparable to observations no rotationally supported discs were found. As strong magnetic braking is responsible for the removal of the angular momentum, this problem is also called the "magnetic braking catastrophe". The results of these simulations stand in contrast to observations which show that discs are present in the earliest stage of protostellar evolution. Here we present results from a number of simulations investigating the role of turbulence in reducing the magnetic braking efficiency and allowing for the formation of protostellar discs.

We now shortly describe the basic simulation setup (see Seifried et al. 2012 for a more detailed description). We simulate the collapse of a 100 M$_\odot$ molecular cloud core which is 0.25 pc in size, threaded by a magnetic field in the z-direction -- the mass-to-flux ratio is $\mu = 2.6$ -- and rotating around the z-axis. Additionally, we add a supersonic turbulence field with a power-law exponent of p = 5/3. The turbulent energy is equal to the rotational energy, i.e. $\beta_{turb}$ = 0.04, corresponding to a turbulent rms-Mach number of $\sim$ 2.5. We performed several simulations with different turbulence seeds to test the dependence of our results on the initial conditions. We find that the results do not depend on the chosen initial turbulence field.

\section{Results}
\label{sec:results}

\begin{figure}[t]
 \includegraphics[width=0.48\linewidth]{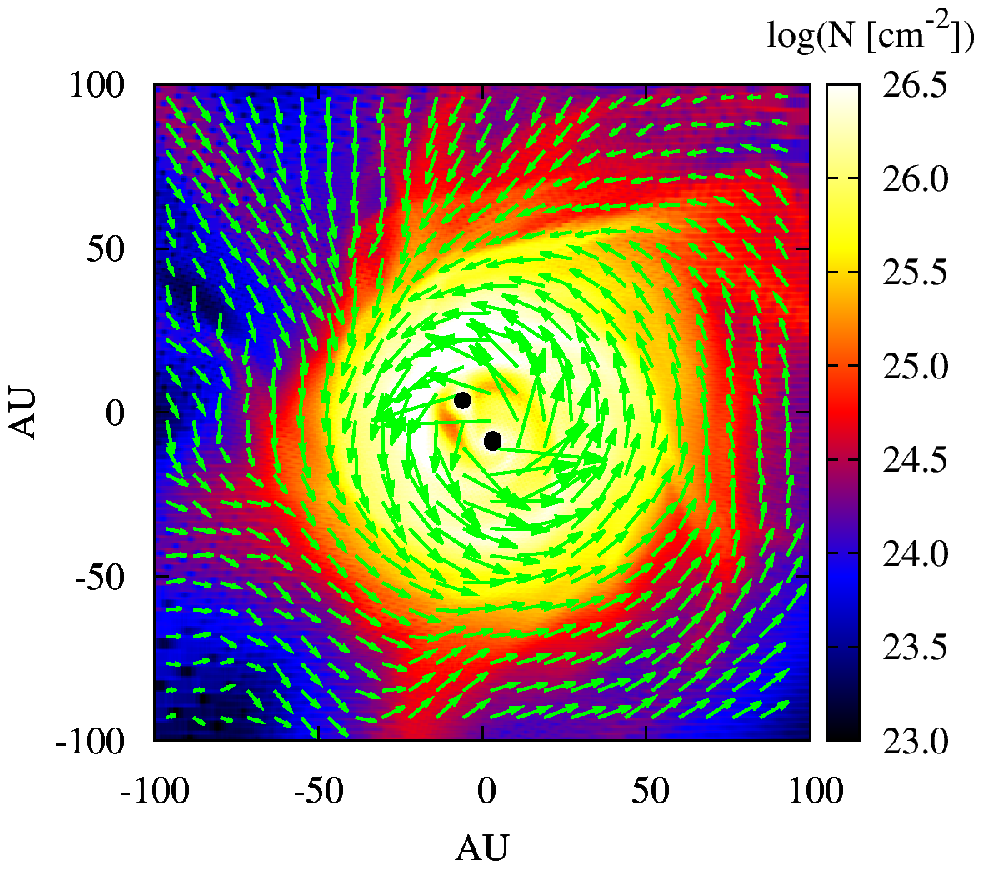}
 \includegraphics[width=0.48\linewidth]{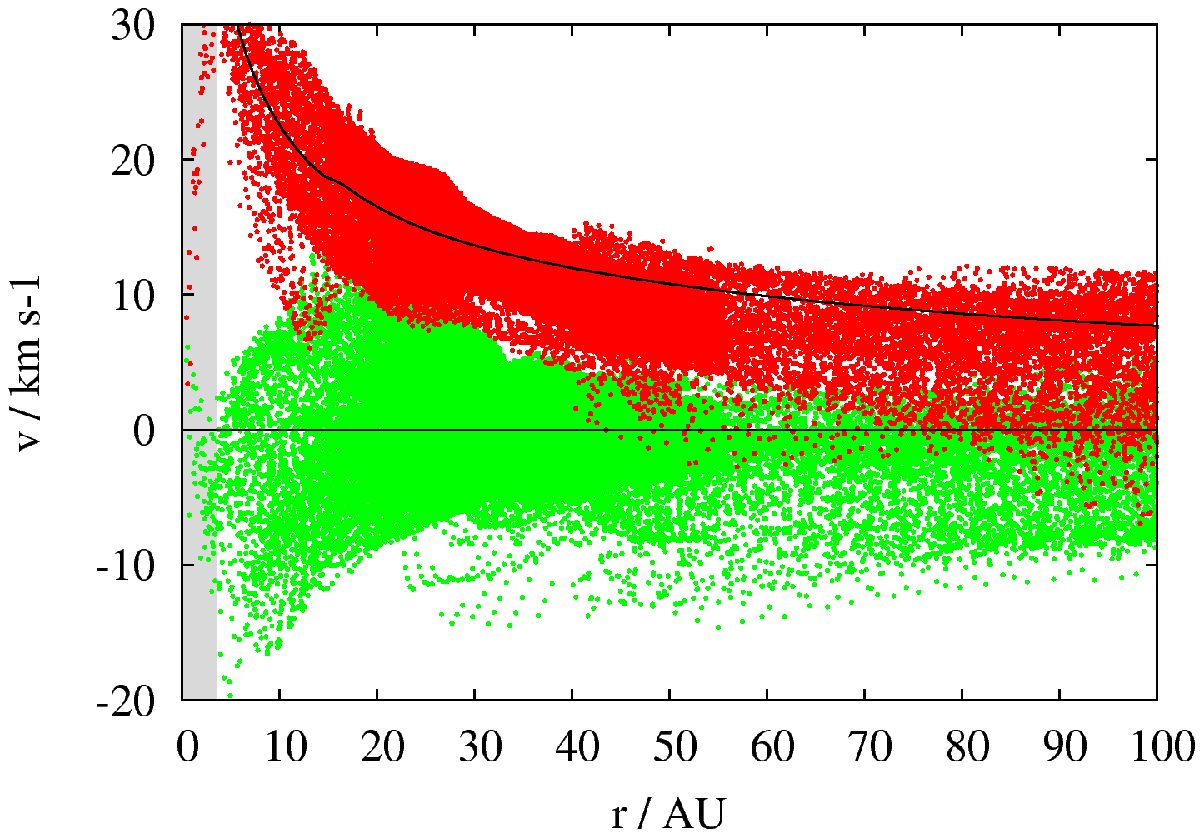}
\caption{Left: Top-on view of a protostellar disc formed in one of the simulations. The velocity structure (green arrows) and the position of the sink particles (black dots) are overplotted on the column density. Right: Velocity structure of the same disc showing the rotating velocity (red dots), the radial velocity (green dots) and the Keplerian velocity (black line).}
\label{fig:disc}
\end{figure}

After a initial collapse phase of about 15 kyr sink particles form in the simulations around which the protostellar discs develop. We follow each simulation for another 15 kyr to analyse the evolution of the discs. In Fig.~\ref{fig:disc} we analyse the velocity structure of one of the discs in detail.
We calculate the rotation and radial velocity in the frame of reference of the disc and show their radial dependence. As can be seen in the right panel of Fig.~\ref{fig:disc}, the rotation scatters around the Keplerian velocity. The radial velocity, in contrast, scatters around 0 and is almost always smaller than the rotation velocity typical for a Keplerian disc. This result is in strong contrast to the results for identical simulations but without initial turbulence~\cite{Seifried11}.

Why, even in the case of such strongly magnetised cores, are Keplerian discs formed? The suppression of Keplerian disc formation in previous studies without turbulence is due to the very efficient magnetic braking which removes angular momentum from the midplane at a very high rate. Hence, in our runs the magnetic braking efficiency has to be reduced significantly.

In a first step we try to estimate the dynamical importance of the magnetic field for the gas dynamics. For this reason we calculate the mass-to-flux ratio in a sphere with a radius of 500 AU around the centre of each disc. We find that $\mu$ is always smaller than 10 which defines the critical value determined in previous simulations below which the formation of Keplerian discs is suppressed. Hence one would expect the magnetic field being strong enough to prevent the formation of Keplerian discs which is clearly not the case as seen in Fig.~\ref{fig:disc}.

\begin{figure}[t]
 \includegraphics[width=0.48\linewidth]{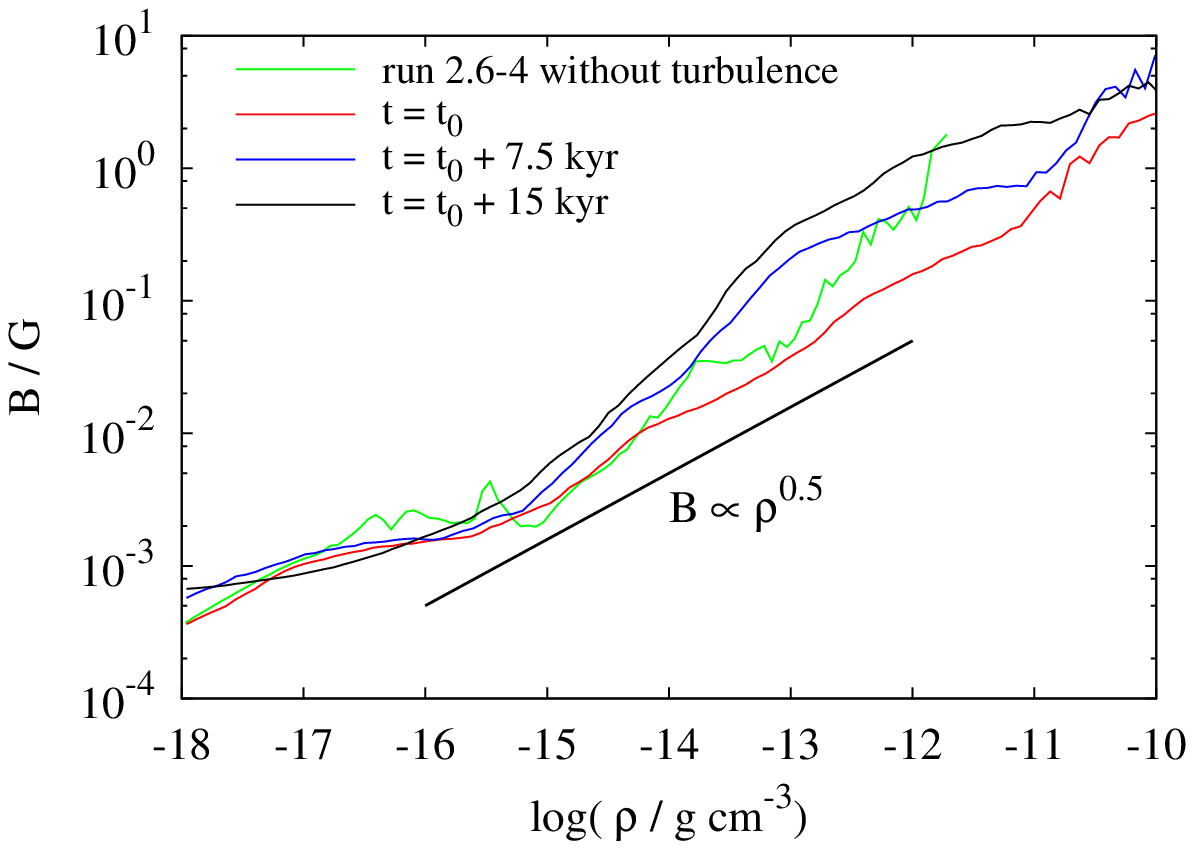}
 \includegraphics[width=0.48\linewidth]{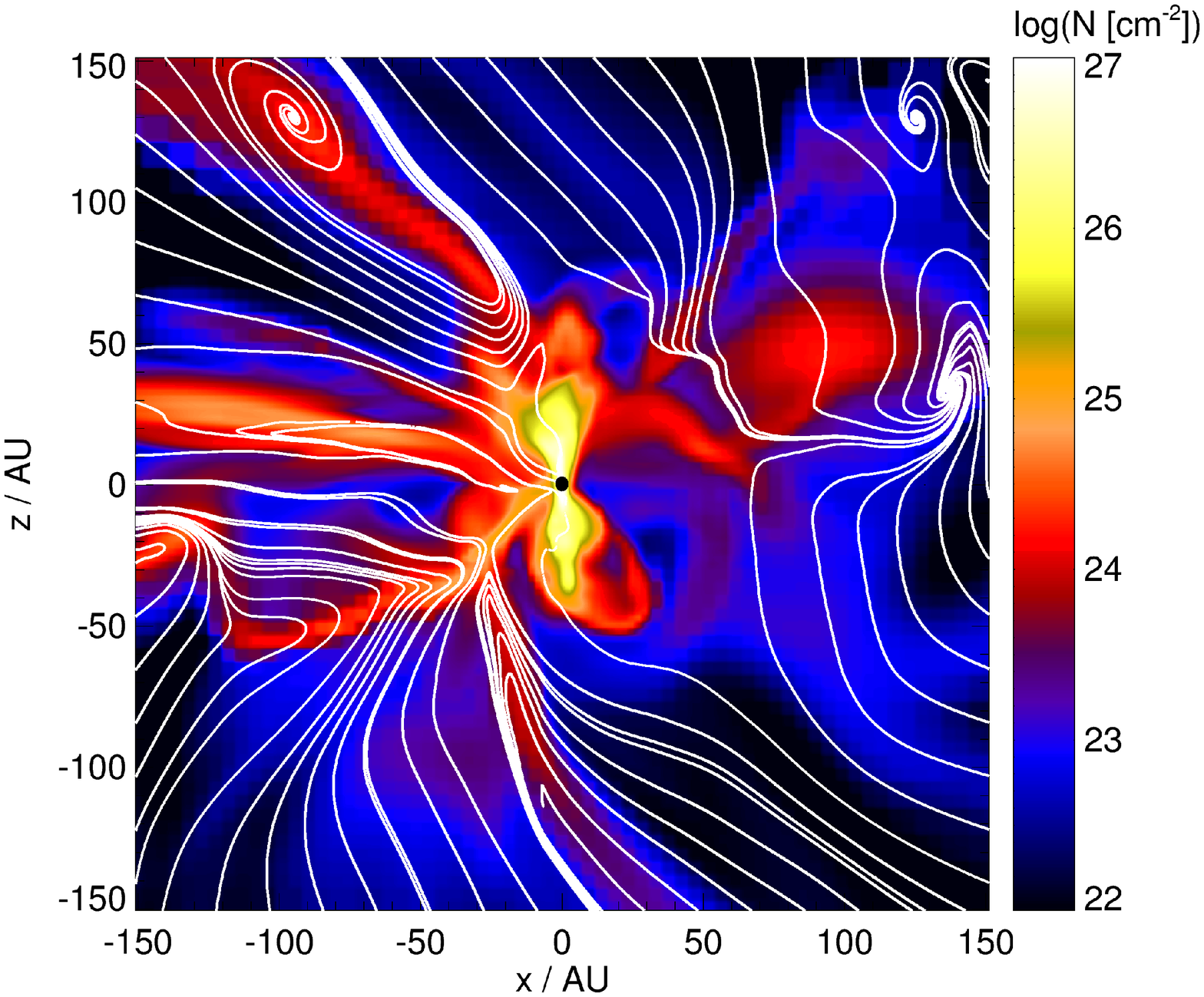}
\caption{Left: Scaling of the magnetic field for different times in one of the turbulence runs as well as in the corresponding
non-turbulent run (green line). Right: Magnetic field structure for the edge-on view of one of the discs formed in the simulations.}
\label{fig:scaling}
\end{figure}

In the left panel of Fig.~\ref{fig:scaling} we plot the scaling of the magnetic field with the density. As can be seen, the observed scaling $B \propto$ $\rho^{0.5}$ is very similar to that of the non-turbulent run (green line). Hence, we argue that in our case no significant turbulent reconnection occurs and that magnetic flux loss is not responsible for the formation of Keplerian discs as proposed recently~\cite{Santos12}.

Considering the magnetic field structure around one of the discs in the right panel of Fig.~\ref{fig:scaling}, one can see that the magnetic field is highly disordered. Hence, it is not surprising that the direction of the mean magnetic field $|B|$ significantly differs from the disc rotation axis which might decrease the magnetic braking efficiency~\cite{Joos12}. However, due to the highly disordered structure the approximation of the magnetic field by a mean magnetic field is at least questionable.

However, this highly disordered magnetic field indicates why the magnetic braking efficiency is reduced strongly. Considering the left panel of Fig.~\ref{fig:disc} it can be seen that in the surroundings of the disc there is a turbulent velocity field with no signs of a coherent rotation structure. Therefore no toroidal magnetic field component can be built up. But as the angular momentum is mainly extracted by toroidal Alfv\'enic waves, it is not surprising that the magnetic braking efficiency is strongly reduced in the environment of the disc despite a low mass-to-flux ratio. Moreover, the disordered magnetic field structure itself impedes the coupling of the fast rotating gas in the inner parts to slowly rotating gas in the outer parts necessary for the magnetic braking mechanism to work efficiently. Hence, the magnetic braking efficiency is reduced due to the lack of a proper toroidal magnetic field component and the highly disordered magnetic field structure in the disc environment. Despite the lack of a coherent rotation structure, locally the inwards angular momentum transport can remain high due to local shear flows driving large angular momentum fluxes. Indeed when comparing the torques exerted by the gas and the magnetic field it can be seen that the gas torque exceeds the magnetic torque by at least a factor of a few. Hence there is a net angular momentum flux inwards~\cite{Seifried12} resulting in the observed build-up of a Keplerian disc. In contrast, for the corresponding non-turbulent run the gas torque is almost perfectly balanced by the (negative) magnetic torque.

\section{Conclusions}

We have observed the formation of rotationally supported discs during the collapse of massive, magnetised and turbulent cloud cores. We attribute this to the turbulent surroundings of the discs which reveal a highly disordered magnetic field and no signs of a coherent rotation structure. In contrast, magnetic flux loss is not able to account for the formation of Keplerian discs as only on the disc scale ($\le$ 100 AU) significant flux loss is observed. However, we emphasise that the magnetic braking efficiency has to be reduced already on larger scales ($\ge$ 500 AU) even before the gas falls onto the disc. This is demonstrated by a corresponding non-turbulent run where the angular momentum is removed largely already outside the disc. Furthermore, non-ideal MHD effects, which have already been shown that they cannot account for the formation of early-type Keplerian discs, are not required as turbulence alone provides a natural and simple mechanism to circumvent the ``magnetic braking catastrophe''. Hence, our work strongly suggests that the magnetic braking problem as reported in numerous papers is more or less a consequence of the highly idealised initial conditions neglecting turbulent motions.

\end{document}